# Marietta Blau and the photographic method of particle detection

Brigitte Strohmaier, University of Vienna, Faculty of Physics – Nuclear Physics, Vienna, Austria

## Abstract

Marietta Blau (1894–1970) was a Vienna born nuclear physicist who eventually became the leading expert in the photographic detection of ionizing particles. She studied physics at the University of Vienna, graduated in 1919. As of 1924, she adapted the technique of photography for the detection of nuclear particles at the Vienna Institute for Radium Research. When after years of painstaking methodical development Blau succeeded in registering tracks of particles, she and her collaborator Hertha Wambacher further refined the technique and by its use discovered the star-shaped tracks of reaction fragments from the interaction of cosmic-ray particles in photographic emulsions, the "disintegration stars", in 1937. Due to her Jewish descent, Blau emigrated in 1938, first to Mexico, later to the U.S. where she had access to scientific research again in 1948. At Brookhaven National Laboratory and Miami University she applied the photographic method to reactions induced by high-energy particles from accelerating machines rather than cosmic rays. When she returned to Vienna in 1960, she supervised the evaluation of photographic plates from CERN experiments.

The article is based on B. Strohmaier, R. Rosner (Eds.), Marietta Blau – Stars of Disintegration. Biography of a Pioneer of Particle Physics, Ariadne Press, Riverside, California (2006).

## 1. The Institute for Radium Research in Vienna

Radioactivity was discovered by Henri Becquerel in 1896 as a property of uranium salts. At the end of 1898 Marie and Pierre Curie, also in Paris, discovered the element radium by extracting its chloride from tailings of the uranium mine in Joachimsthal (Bohemia), supplied by the government of the Austro-Hungarian Empire. In Vienna, radioactivity was studied by Stefan Meyer (Fig. 1) and Egon Ritter von Schweidler at the physics institute of the University of Vienna as early as 1899. Also the Imperial Academy of Sciences turned its interest to radioactive substances, founding a commission for their investigation, and had radium extracted from ten tons of tailings from the Joachimsthal uranium production, thereby owning the largest quantity of radium in the world. But the equipment of the physics institute, at that time located in a rental apartment in the 9$^{th}$ district, was insufficient to adequately take advantage of this scientific treasure. It was Karl Kupelwieser, an advocate with connections to both mining industries and science, who made a donation to the Academy of Sciences in 1908 to have an institute constructed and equipped for the physical investigation of radium. The building lot should be chosen in a place where new university institutes for physics and chemistry could be erected, too, and the government should also operate the Institute for Radium Research of the Imperial Academy of Sciences. Stefan Meyer was entrusted with planning both the building and the equipment and was, therefore, considered its creator. The institute was dedicated in October 1910 (Fig. 2), the first Radium Institute in the world and example for similar institutions built later on in Paris, St. Petersburg and Warsaw. The staff of the Vienna Radium Institute (VRI) consisted of employees partly of the Academy of Sciences and partly of the University of Vienna. The first director, Franz Serafin Exner, did not actively engage in the scientific activities, hence Stefan Meyer, formally holding the position of substitute director from 1910 to 1920, was *de facto* head of the institute, which he became *de iure* in 1920.

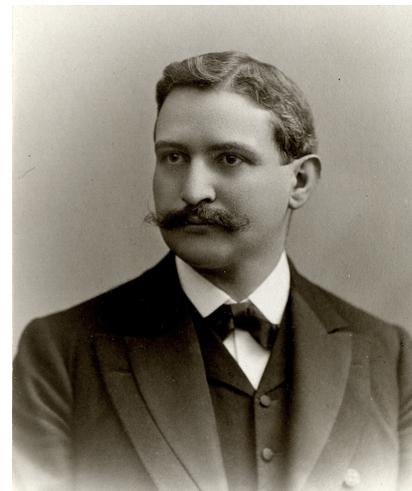

Fig. 1: Stefan Meyer (Austrian Central Library for Physics)

Outstanding scientific success was achieved by Viktor Hess who investigated the radioactivity of the atmosphere, in particular by electrometer measurements of the ionization (number of ions per unit volume) as function of altitude on a number of balloon rides. The result, an increase of ionization with increasing altitude, was explained by an extra-terrestric radiation entering the Earth's atmosphere.[1] Hess's discovery of "cosmic rays" was awarded (half) the Nobel Prize in Physics in 1936.

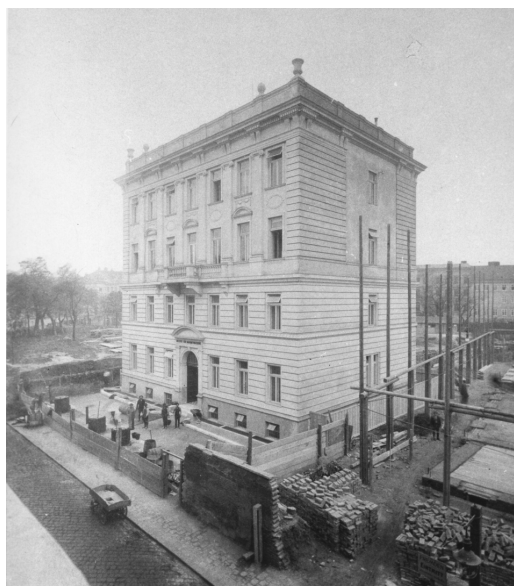

Fig. 2: Institute for Radium Research, 1910 (Austr. Acad. Sci., Photo Archive A-0144-D)

Georg de Hevesy and Friedrich Paneth, on the other hand, did research in the field of radiochemistry and put to use the inseparability of radioactive and stable variants of the same element. The "radioelement", once added in defined proportion, undergoes physical and chemical processes in constant relation to the stable element. As incomparably smaller amounts of radioactive than of inactive substances can be measured due to their activity, traces of a radioelement added to the stable element can serve for its qualitative and quantitative detection. Hevesy and Paneth first published the use of radioelements as indicators in analytical chemistry in 1913.[2] For applying the method of radioactive indicators in investigating chemical processes, Hevesy won the Nobel Prize in Chemistry in 1943.

A new field of study opened up when scientists directed particles of radioactive decay on stable nuclei, inspired by Rutherford's publication of an observed "transmutation" of nitrogen to oxygen.[3] This process had been induced by polonium α-particles passing through nitrogen gas and was later identified as the nuclear reaction $^{14}N(\alpha,p)^{17}O$. At the VRI, the investigation of such "atomic disintegrations" was initiated by Hans Pettersson, an oceanographer from Göteborg, Sweden, who had come to the Radium Institute to measure the radioactivity of samples of deep-sea sediments. He was impressed by the excellent equipment and liberty of research at the VRI and stayed for several years, dedicating his work to nuclear reactions.

In order to prove an (α,p) reaction its fragments, i. e., the emitted protons, had to be detected. At that time, this could be accomplished only by the scintillation method which relies on the phenomenon that ionizing radiation causes small light flashes on sphalerite (ZnS) screens. Their observation with moderately powered microscopes required that the eyes adapt to the darkness of the laboratory in order to detect the weak luminosity, a procedure which was fatiguing for the observers and hence subject to errors. Therefore, it was considered desirable to improve the methods for detecting fast protons. The Institute's scientists were energized by Pettersson not only to develop alternative detection methods to replace the scintillation method, but also to investigate thoroughly the scintillation process itself in order to try to overcome the deficiencies mentioned. This task was assigned to Berta Karlik (born 1904) as doctoral thesis. Alternative methods of particle counting comprised various electrical techniques, like modifications of ionization chambers, and Wilson's cloud chamber (not in use at the VRI until 1926/27), and the photographic method whose employment for detection of the particles emitted in nuclear reactions was entrusted on Marietta Blau in 1924.

## 2. The photographic method of detecting electromagnetic and particle radiation

In a photographic emulsion, small crystals of silver bromide (AgBr) are distributed in a transparent matrix of gelatin. The emulsion is coated on a thin glass plate. In the crystals, the elements are arranged as ions in a lattice ($Ag^+$, $Br^-$). All silver halides are sensitive to electromagnetic radiation and ionizing particles in that electrons produced by the radiation neutralize silver ions to form atoms which accumulate on silver development seeds existing in the emulsion. The resulting silver clusters form a latent image which is stable over long time and can be turned into a visible image by developing. The developer, a reducing agent, renders electrons to the silver ions which are reduced to metallic (visible) silver, preferably at the clusters of the latent image produced by exposure to electromagnetic or particle radiation. In a fixing bath the unexposed parts of the layer are washed out.

In 1895, x-rays were discovered. They are electromagnetic radiation of higher energy than visible light, and their photography was used for imaging in medicine and crystallography from the beginning. Guido Holzknecht (1872–1931), a medical doctor at the Vienna General Hospital, became head of the newly installed Central X-ray Institute there. The use of x-ray systems for diagnostics was the basis of his international reputation as medical radiologist.

Radioactivity had been discovered thanks to the photographic effect of the radiation of uranium salts. Hence it was straightforward to apply photographic emulsions to register the tracks of α-particles from radioactive decay. Wilhelm Michl studied the photographic effect of α-particles at the VRI in 1914[4] and found that they blacken the emulsion at discrete places, i. e., produce sequences of black dots (silver grains). In order to register the complete tracks, the preparation was located such that the α-particles (from preparation p; Fig. 3) entered the emulsion at a very small angle to the surface (grazing incidence). Michl understood that the track length is a measure for the energy of the particle which loses its energy in successive collisions with emulsion nuclei and, therefore, also investigated how the geometric measures of the emulsion changed during developing. After Michl had died in World War I, the method was not further pursued.

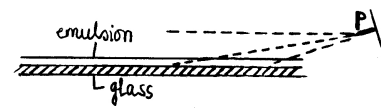

Fig. 3: Paths of α-particles incident on a photographic plate

## 3. Marietta Blau's family and education

Marietta ("Etta") Blau was born on April 29, 1894, as third child and only daughter to the lawyer Dr. Markus Blau and his wife Florentine, née Goldenzweig, an upper middle-class Jewish couple. The family lived in Vienna's second district, which had been the location of the Jewish ghetto in the seventeenth century. Markus Blau came from Deutschkreutz, a village in West Hungary with a very religious Jewish community that strictly observed all the traditional regulations. But like many Jewish families who migrated to the rapidly developing capital of the Habsburg Empire in the nineteenth century, he loosened his ties to Jewish tradition.

When the first-born son, Fritz, died in 1895, this left Marietta with an elder brother, Otto, and a younger one, Ludwig. After attending the five-grade practice school of the training college for teachers in Hegelgasse 12 (1st district), Marietta went to the private high school for girls run by the Association for the Extended Education of Women (*Verein für erweiterte Frauenbildung*) starting in 1905. This school, founded in 1892 and located at that time at Hegelgasse 19, was a novel venture in the education of women in Austria and the first to make it possible for girls to obtain the general certificate of education (*Matura*). This certificate was (and still is) conferred after a rigorous examination in the most important subjects and entitles a student to enroll at any university without further examinations. In the beginning, however, the girls had to take the exams at the renowned Academic High School for boys. In 1910 the high school for girls of the *Verein für erweiterte Frauenbildung* became a public school and moved to Rahlgasse 4 (6th district). Marietta Blau obtained the *Matura* with distinction in July 1914. In fall of 1914, she enrolled at the University of Vienna as a regular student of physics and mathematics. With a doctoral thesis on a radiological topic, *The absorption of divergent $\gamma$-rays*,[B1] Blau obtained her PhD in March 1919. Subsequently, she spent several months as an observer with Guido Holzknecht at the Central X-ray Institute of the Vienna General Hospital.

At that time, far-reaching changes were taking place in Europe, in that both the Austro-Hungarian and the German Empire collapsed. The infancy of the newly constituted Austrian Republic was accompanied by widespread starvation and considerable political unrest.

In the second half of 1921, Marietta Blau was employed as a physicist at the x-ray tube factory of Fürstenau, Eppens & Co. in Berlin, where she conducted investigations in electrical engineering and spectral analysis. At the beginning of 1922, she changed to a position as assistant professor at the Institute for the Physical Bases of Medicine at the University of Frankfurt/Main. She was in charge of scientific investigations for the electrotechnical and electromedical industry, as well as conducting theoretical and practical x-ray training for doctors and instructing doctoral students. Together with Kamillo Altenburger, she published papers on the absorption and theory of the effect of x-rays.[B2,B3] When her mother, widowed since 1919, fell ill in the autumn of 1923, Blau resigned her position and returned to Vienna. Professionally, she exchanged a paid job for unpaid work at the VRI. As a result, she had to rely on sustenance by her family since taking care of her mother became her first priority and was to be a determining factor in her life over many decades.

At the VRI, the percentage of women was especially high (over one third). This has been attributed to the "magic spell" of radioactivity and to the success of Marie Curie at the Paris Radium Institute, but a more plausible explanation is the fact that with the discovery of radioactivity a new field of physics opened up just at the time when women – thanks to their recent admission to science studies at universities – could for the first time acquire the necessary qualifications for physics research and hence have access to a scientific field not yet in the hands of the male establishment. Most of them were members of the *Verband der Akademikerinnen Österreichs*, the Austrian branch of the International Federation of University Women. At the VRI between 1910 and 1938, a large portion of the scientists were doctoral students or graduates who worked at the Radium Institute without pay which required adequate financial support from their families. On a positive note, this custom provided the opportunity, particularly for women, to gain a foothold in physics research.

The director of the VRI, Stefan Meyer (1872–1949), was a father figure to many of his collaborators and fostered a family-like relationship among the staff at the Radium Institute. His intense personal involvement with his co-workers and their needs proved to be a positive in-fluence on the scientific work at the Institute; the gratitude repeatedly expressed towards him in the Institute's numerous publications was well founded.

The cordial atmosphere characteristic of the Radium Institute at that time contrasted with the situation in many other parts of the university. The long tradition of anti-Semitism at Austrian universities gained considerable influence in the academic world after World War I. Throughout the 1920s, there were numerous instances of anti-Semitic students assaulting Jewish professors and forcing Jewish students to leave lecture halls. Not only the German nationalist student organizations (the predecessors of the Nazi organizations), but also Catholic student organizations demanded that restrictions be imposed on the admission of Jewish students. It is not surprising that many Jewish students and university staff, among them Marietta Blau, no longer felt secure when ideas of this kind were becoming more and more prevalent. Blau herself was described in 1926 as "a little Jewess, always looking out for the next pogrom."

## 4. Marietta Blau's development of the photographic method

Analogous to the reaction $^{14}N(\alpha,p)^{17}O$ observed by Rutherford, the susceptibility to disintegration (disintegrability or *Zertrümmerbarkeit*) of other elements was investigated in laboratories all over Europe. Blau was to tackle the investigation whether the emitted fast protons "could be made amenable to objective observation through their photographic effect." (H-rays, H-particles, or H-nuclei were the terms used then for fast protons.) It was clear that if photographs of particle paths could be taken, the tracks would remain stored on the plates and enable subsequent analysis, in contrast to the quickly vanishing scintillations.

This question may have been appealing to Marietta Blau (Fig. 4) since she had studied the systematics of the photographic effects of x-rays and compared them to those of visible light at *the Institut für Physikalische Grundlagen der Medizin* in Frankfurt/Main in 1922.[B2] The finding was that a single hit of an x-ray photon renders a grain of silver after development, whereas the threshold value in the gradation curves of visible light is explained by the assumption that a silver-bromide grain or a molecule complex needs to be hit by light photons more than once in order to be influenced effectively, as if it were excited in a multi-step process to a final state yielding a silver grain when developed.

At the Vienna Radium Institute, Blau's goal as of 1924 was to register tracks of fast protons as series of dots in photographic plates. Her means of getting there were variations of the following parameters:

- Grain size, i. e., average size of AgBr crystals in the emulsion;
- Layer thickness of the emulsion;
- Development conditions, in particular duration and temperature of development as well as composition and concentration of the developer;
- Pre-treatment of the photographic emulsion to influence the sensitivity as was known from spectral sensitization (covering the crystals with coats of dye molecules) in photography of visible light.

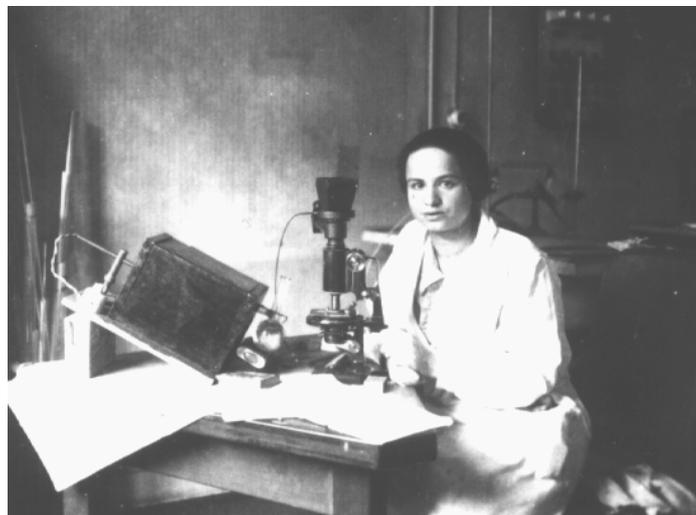

Fig. 4: Marietta Blau around 1927 (Archive Agnes Rodhe née Pettersson)

For the experimental setup Blau had to take care of the following items:

- Construction of a source of protons with higher yield than (α,p)-reactions: Blau turned to using hydrogen-containing substances from which protons were ejected by incident α-particles in nuclear collisions ("natural H-rays"). In practice, α-radiating $^{210}$Po was deposited electrolytically on small sheets of gold or platinum on which layers of paraffin were placed where the α-particles collided with the H-nuclei, thereby ejecting them. In order to prevent α-particles from reaching the photographic plates, the paraffin was covered with an absorption foil of copper, which does not keep the protons from leaving the paraffin.
- Grazing incidence on the plates: The protons from the source described above were selected by an aperture to achieve an angle of incidence of about 30°. (In Fig. 3, p now represents the arrangement of preparation, paraffin and absorber; the particle paths are those of protons.)
- Choice of photographic plates: Blau experimented with various types and pointed to factors that determine the suitability of emulsions for the detection of proton tracks: fine grain (i.e., small size of the AgBr crystals) and low sensitivity to competing radiation such as β- and γ-rays emitted by the radioactive preparations as well as to light from luminescence and the laboratory environment.
- Test experiments: Replacing paraffin with soot (almost pure carbon) yielded a significantly less silver-grain formation; thus, Blau made sure that the majority of the tracks obtained when paraffin (high hydrogen content) was used indeed came from the H-rays.

After long and tedious studies, Blau succeeded in detecting natural H-particles.[B5,B6,B8] In the photographic layer, they act much the same as α-particles: They produce rows of black dots which correspond to developed grains. (In Fig. 5, a row of fifteen grains[B8] is seen in the right half above the dark stain; its actual length is 86 μm.) The length of a row is a measure for the range of the particles in the emulsion, i. e., the mean distance the particle travels until it is stopped due to successive energy losses, and hence its total energy. In order to obtain the particle energy, therefore, the total track length had to be recorded which due to the thinness of the emulsion layers (≤ 200 μm) was only achieved if the protons passed the emulsion approximately parallel to the layer surface. Using thicker emulsions required appropriate development conditions to enable a homogeneous development of the complete layer.

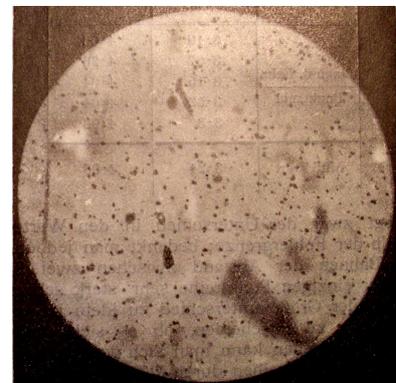

Fig. 5: Track of a proton produced by a polonium α-particle colliding with hydrogen (B8)

Turning to the original goal, Blau carried out experiments to detect protons from a nuclear reaction, namely, $^{27}$Al(α,p)$^{30}$Si that had already been observed by other methods. The polonium on the flat metal backing was now covered with aluminum foil instead of paraffin layers. She found rows of silver grains in the expected direction, verifying the disintegrability of aluminum by α-particles.[B6,B9]

Blau evaluated thousands of tracks of α- and H-rays with regard to track length and number of blackened grains with the microscope in order to obtain quantitative data on the photographic effect of these particles. By comparison of photographic with ionization-chamber detection, it turned out that the protons had been registered almost quantitatively with a new fine-grain emulsion. Their energy spectrum was derived as diagram of track length vs. number of occurrence.[B13]

Blau found that the number of developed grains in proton tracks registered with the fine-grain plates was twice that in coarse-grain Agfa dental x-ray film.[B13] She tested various types of photographic plates commercially available, but she also corresponded with the head of the scientific central laboratory of the photographic department of Agfa, and had emulsions made according to her specifications. It was common that scientists had special products made on their request. In 1932, Blau received the International Senior Fellowship of the Austrian University Women which she used to conduct studies in crystal physics in Göttingen. On her way there she stopped at Leipzig for a visit to Agfa. She spent two days talking to an Agfa employee who worked on the effect of radioactive and x-rays on the photographic layer.

Comparing particle types,[B23] the ionizing energy spent per path unit for α-particles is four times that for protons at the same velocity. Moreover, the specific ionizing power for corpuscular radiation is inverse proportional to velocity. Therefore, α-particle tracks are well defined due to high density of developed grains. The distance of subsequent dots blackened by α-particles is practically the same for all track lengths, that is, for all α-energies all AgBr grains lying on the particle path are blackened. For natural H-rays, the grain distance decreases with decreasing track length, corresponding to a high density of black dots in the tracks of low-energy protons. The onsets of long proton tracks (high energy) may not show up in the emulsion, and even for the visibility of slow protons, not only the number of formed silver seeds is essential, but in particular their localization on the grain surface. Again, that suitable surface sites are affected is less likely for protons due to the smaller cross section, whereas for α-particles the number of freed silver ions is sufficiently large that the grain will be developable. The optical density is proportional to the number of incident particles. (A grain will be capable of development with ≥ 300 silver atoms.)[B23]

Blau understood that the suitability of the photographic method as a technique for registering fast protons depended on whether the total number of tracks, the complete length of the tracks, and a large number of black dots per cm (high density of developed grains) could be made visible in a reproducible manner without fogging.[B23]

She discussed general questions of photographic processes, in particular the mechanism of the photographic effect itself, starting from the observation that particles of radioactive decay render a grain developable by one hit, like x-rays do, rather than by a multi-step process.[e.g. B10] (In contrast to visible light which neutralizes interstitial silver ions, incident x-rays or charged particles affect these in regular lattice positions.)

In view of reliable use of photographic plates for quantitative radioactivity measurements, Blau started studying in detail the presumed fading of the latent image (i.e., whether and how fast the invisible image consisting of activated silver ions vanishes) after exposure to α-particles.[B16] Information about fading is necessary if the photographic plates are stored for some time between exposure and development, and especially for low-level measurements in which it is desirable to choose long exposure times to register a significant number of events on a single plate. She found that the optical density produced by radiation decreases gradually, as much as 20% in sixty days. The effect is explained as a recrystallization process (healing) in the silver-bromide grains in which the radiation had caused lattice defects.[B16]

Marietta Blau published eight papers[B5,B6,B8–B10,B13,B15,B16] on her methodical studies of the photographic method as single author between 1925 and 1931. Her attempts to acquire the right of teaching at universities (*Habilitation*) were turned down by doubled discrimination: "A woman and a Jew – that is too much!"

## 5. Hertha Wambacher – pre-treatment of emulsions and neutron detection

In 1928, Blau began collaborating with Hertha Wambacher (1903–1950; Fig. 6), whose doctoral thesis Blau supervised.

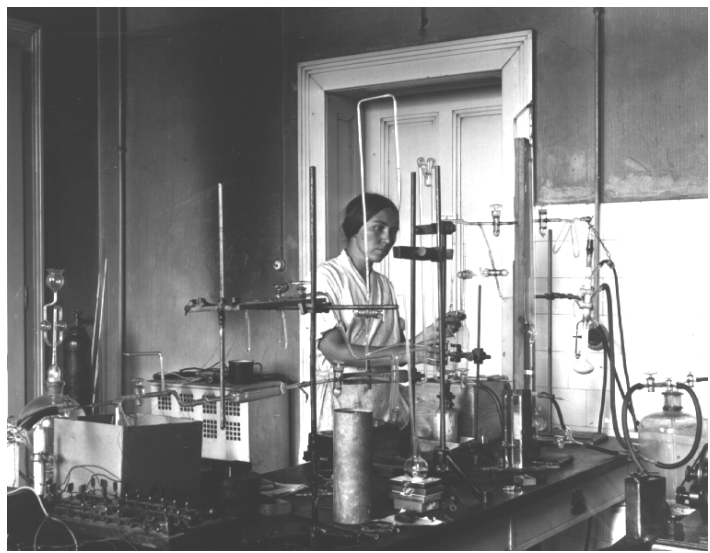

Fig. 6: Hertha Wambacher (Archive R. & L. Sexl)

Hertha Wambacher, a manufacturer's daughter nine years Blau's junior, had also attended the high school for girls of the Association for Women's Extended Education. These two women seem to have been complete opposites, both in appearance and character. Perhaps the fact that they were alumnae of the same school brought them together just as their different characters attracted and complemented one another.

Wambacher investigated how the impregnation of photographic plates with certain chemicals affected their response to particle radiation.[5] As recognizing and measuring the rows of developed grains corresponding to particle tracks was impeded by the presence of black dots caused by light, β-, and γ-radiation, this pre-treatment aimed at desensitizing the emulsion to such background radiation without changing its sensitivity to α- or proton rays. The background was suppressed indeed by bathing the plates in pinacryptol yellow, a dye, before exposure, but in addition, α-particle tracks became richer in contrast and grain density, and fast protons which had not been detectable at all, yielded observable tracks after this treatment. The desensitizer (with regard to visible light) sensitized the plates for high-energy protons. Wambacher's thesis was approved in December 1930; she graduated in May 1932. Thereafter, Blau and Wambacher worked together for six years on methodical investigations of the photographic method. They examined a large number of substances for impregnating photographic plates.[B23,B27,B28,B32] At first, the effect of desensitizers, various types of dyes, could not be explained, even though they were used empirically. Blau and Wambacher, among others, attempted to provide an explanation:[B25,B29] The effect of the dye (as sensitizer for particle radiation) is not related to its desensitizing power for visible light. Due to the adsorption of the dye, the surface conditions of the AgBr grains are changed and processes called seed isolation take place. The additional seeds render an increased number of grains developable in the latent image. For the explanation of the phenomenon of desensitization of emulsion plates by certain dyes Marietta Blau, along with Hertha Wambacher, was awarded the medal of the Photographic Society of Vienna.

In 1932 James Chadwick discovered the neutron,[6] an electrically neutral particle of mass slightly larger than the proton mass. Being uncharged, neutrons do not interact with electric fields and hence cannot be detected directly, but via protons resulting from collisions of neutrons with hydrogen nuclei. When neutrons from the reaction $^9Be(\alpha,n)^{12}O$,[7] hit photographic plates that were again pretreated with pinacryptol yellow, they caused rows of dots on the photographic plates because the hydrogen content of the emulsion itself was sufficient to produce the recoil protons.[B18,B19] From the lengths of the tracks, proton energies and consequently the energies of the neutrons could be estimated. It turned out that these energies (up to 9 MeV) were higher than assumed on the basis of other investigations at that time.[B19] Blau continued her experiments on neutron detection via recoil protons when after her visit to Göttingen she worked for a couple of months in 1933 at the Paris Radium Institute upon invitation from Marie Curie.[B22] For several years, the photographic method remained the only one able to detect the fastest of these neutrons. Experiments to determine the energy spectrum of protons emitted in the reaction $^{27}\text{Al}(\alpha,\text{p})^{30}\text{Si}$ were started in Paris[B22] and carried on with Wambacher in Vienna.[B24]

"For their study of the photographic effects of α-radiation, protons and neutrons," Blau and Wambacher were awarded the Ignaz L. Lieben Prize in 1937. This prize had been established through a foundation from the bequest of Ignaz L. Lieben, a Jewish banker who died in 1862, and was awarded through the Vienna Academy of Sciences. The prestige attached to the award of the Lieben Prize was considered more important than its financial value.

## 6. Cosmic rays: fast protons and spallation stars

As soon as the photographic method was established as a tool for the detection of charged particles, another field of application opened up, namely the exploration of cosmic rays. Marietta Blau and her collaborator Hertha Wambacher turned to this field in 1932,[B35] twenty years after the discovery of cosmic rays. At this time, conclusions on the nature of cosmic rays had been derived from ionization measurements at stations at different latitudes and from combined balloon and mountain data. Both the geographical distribution and the intensity as function of altitude were consistent with the assumption of charged particles as constituents of cosmic rays, as was the Bothe-Kolhörster experiment.[8] But there was no way of reconciling the data with the hypothesis that primary cosmic rays consist of photons.[9]

The fact that photographic plates accumulate events over long periods of time made them particularly suited to detect the rare cosmic-ray hits. The knowledge of fading of the latent image turned out to be most valuable in cosmic-ray research with exposures lasting several months. In order to register as large a fraction of ionization tracks of high-energy particles as possible, increased thickness of the emulsion layers in the photographic plates was considered indispensable. To this end, Blau directly contacted Ilford Ltd. in England. New developing techniques were needed for the increased layer thickness.

In the 1930s, flexible emulsion carriers made from nitro or acetate cellulose (roll film) were in common use. In scientific photography, however, for which form stability, planeness and resistance against environmental influence are of importance, photographic plates (plane film) were still preferred.

In 1937 Blau and Wambacher approached Viktor Hess, the discoverer of cosmic rays, who in 1936 had been awarded the Nobel Prize in Physics for his pioneering work. In his observatory at Hafelekar, a 2300-meter mountain north of Innsbruck, the intensity of cosmic rays was constantly recorded, initially with ionization chambers. Upon their request to Hess, Blau and Wambacher were allowed to expose photographic plates at the cosmic-ray observatory.

On the plates (Ilford New Halftone, 70 μm emulsion, no pinacryptol-yellow bath) exposed for four months at Hafelekar, many long tracks were found and assumed to be those of protons, either from recoil by neutrons or from atomic-disintegration processes, but not primary particles themselves, as no strong preference of vertical incidence was observed. Numerous tracks corresponded to ranges in air of over one and up to twelve meters.[B35,B38] Due to these large ranges, most tracks did not end in the emulsion, hence the observed lengths did not equal the ranges (energies) of the particles. Therefore, these were determined from the distances of developed grains in the tracks. To enable such an evaluation, Blau and Wambacher analyzed a large number of tracks and derived a (preliminary) relation between grain distance and particle energy, ending up with an average particle energy of 12 MeV.

Even more importantly, a new pattern of tracks was discovered, namely that of several reaction products starting where a cosmic-ray induced nuclear reaction had taken place. Due to the starlike shape of these tracks (Fig. 7), they were called disintegration stars (*Zertrümmerungssterne*). From a pronounced center, up to twelve tracks proceed, varying in grain density and length. For several of sixty registered stars of "multiple disintegration" (spallation), Blau and Wambacher estimated the total energy of emitted particles to be several hundreds of MeV.

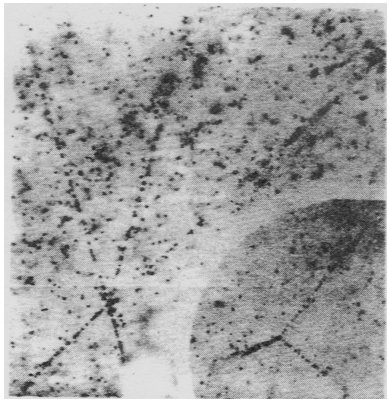

Fig. 7: Two disintegration stars (enlargement 300×) (B38)

Wambacher was the ambitious one who urged to publish the results[B36,B38] as soon as possible in order to keep others from getting in front of them. In September 1937 Blau made a request of Friedrich Paneth who had worked at the Radium Institute for several years before 1920 and was now professor at the Imperial College in London, whether they might include packets of photographic plates on balloon flights into the stratosphere in order to register particle tracks of cosmic rays in the emulsions. What followed was a lively correspondence on feasibility, achievable altitude, weight and wrapping of the photographic-plate packets, and optimization of the emulsions. Blau and Wambacher even received a grant from the Academy of Sciences enabling balloon flights with the emulsions. All this was interrupted, however, by the political events in Austria in 1938.

## 7. Marietta Blau's emigration: Norway, Mexico, U.S.A.

The discovery of the disintegration stars was met with great interest in scientific circles.[10] The theory of nuclear forces was just at the very beginning of its development. Werner Heisenberg had worked out a theory about cosmic-ray particles colliding with atomic nuclei leading to multi-particle emission.[11,12] Blau and Wambacher's detection of such events now provided experimental evidence for his ideas and furthered the discussion of such processes.

Last but not least, Albert Einstein was impressed by Marietta Blau's successful method of studying cosmic rays by photographic means. On February 14, 1938, four weeks before Austria's annexation, Einstein intervened to find a post for Blau in Mexico. He mentioned the modest means her research method required, which made her research particularly suited to a country like Mexico. Einstein was aware of the fact that Blau was being ousted in Vienna as a Jew for political reasons. And indeed, despite the great scientific success Blau and Wambacher had in 1937, the relationship between the two seems to have deteriorated rapidly as a result of political developments. The same was reported by Ellen Gleditsch, a professor of inorganic chemistry at Oslo University who had visited the VRI in 1937/38 and witnessed the Nazis' attitudes there. To rescue Blau from the kind of treatment she was receiving, Gleditsch invited her to Oslo for spring and summer of 1938.

### 7.1. Norway

Blau left Vienna on March 12, 1938, the evening before the Nazi invasion of Austria, one of the last Austrians to pass the German border. When meeting the German troops on her trip, she realized that probably she would not be able to return from Oslo to Vienna. In April 1938, Einstein made an unsuccessful effort to find a position for her in the U.S., but his attempts to procure a professorship in Mexico for Blau brought the desired result. The head of the department of technical studies, Juan de Dios Batiz, offered Marietta Blau a position as a professor for advanced students at the *Instituto Politécnico Nacional* (IPN) in Mexico City. The foundation of this institute had been an important step in the political and cultural reforms of president Lázaro Cárdenas (elected in 1934). These reforms were received with great interest among the progressive intellectuals and may have been the reason Einstein considered Mexico a suitable place for Blau to kindle scientific research.

Blau now had to get her permit of residence in Norway extended and obtain a German passport (with the stamp "J" [Jew]) because her Austrian passport was no longer valid, and an immigration visa for Mexico. She also had to get her mother out of Austria, as she had decided to take her along and to care for her in Mexico. After nerve-wrecking delays, Blau met her mother in London in the first half of October 1938; together they arrived in Mexico by ship at the beginning of November.

## 7.2. Mexico

Blau started teaching at the *Escuela Superior de Ingeniería Mecánica y Eléctrica* (ESIME) of the IPN on January 1, 1939. Right from the start, she was disappointed with conditions at the institute: Neither could she develop photographic plates there, nor did she have a microscope. Even for her classes – she was involved with the program for graduate studies – the most basic equipment was missing. She built a Geiger counter herself, and to acquire data, she got her students to take photographic plates for registering cosmic rays with them on mountain trips.

Mexico was the only country that formally protested before the League of Nations in Geneva in March 1938 against Nazi Germany's annexation of Austria and that never acknowledged the validity of the *Anschluss*. Exile in Mexico saved Blau from the Shoah, but she had to cope with conditions that were highly unsatisfactory: She felt the resistance of some of the professors of the IPN who were of German origin and at that time supportive of the Nazis. At the ESIME, engineers with their focus on industrial progress undermined the pursuit of pure science. Moreover, the academic atmosphere was one of male dominance which was characteristic of Mexico at that time. Blau was literally marginalized among the teaching staff (Fig. 8).

Blau was invited to join the newly founded commission for proposing and coordinating scientific and technological research (*Comisión Impulsora y Coordinadora de la Investigación Científica*, CICIC). Heading its radioactivity laboratory, she studied the radioactivity of minerals and springs in several parts of the country, particularly in the state of Chihuahua, where minerals containing uranium were found in non-working mines.[B45,B46] Blau reported on her investigations not only in the Mexican journal *Ciencia* but also in the *Yearbook of the American Philosophical Society*.

In 1941, Blau gave a series of lectures at a provincial Mexican university in Morelia and was offered a teaching position there. This appeared attractive to her, as she would have been the only physicist in Morelia, far from all competitive bickering, and with a physics laboratory that the rector had just purchased and which was sitting there in boxes as there was no one capable of installing it properly. When she went to Mexico City to settle the matter in the ministry, she was offered a permanent teaching position at the IPN, which ensured a supposedly secure salary but excluded any chance of pursuing scientific research. Friends and also superi-ors whom she consulted urged her strongly to accept the post and viewed it as the only safe job she could ever hope for in Mexico. Meanwhile, the apparatus in Morelia had disappeared from the university and reappeared in a pawnshop. Time and again she was not paid even her minimal salary from the IPN of approximately hundred dollars per month; the "safe payment" required painstaking efforts on her side, even Einstein's intervention, and in the end became uncertain again due to a law that forbade payment of overdue salaries.

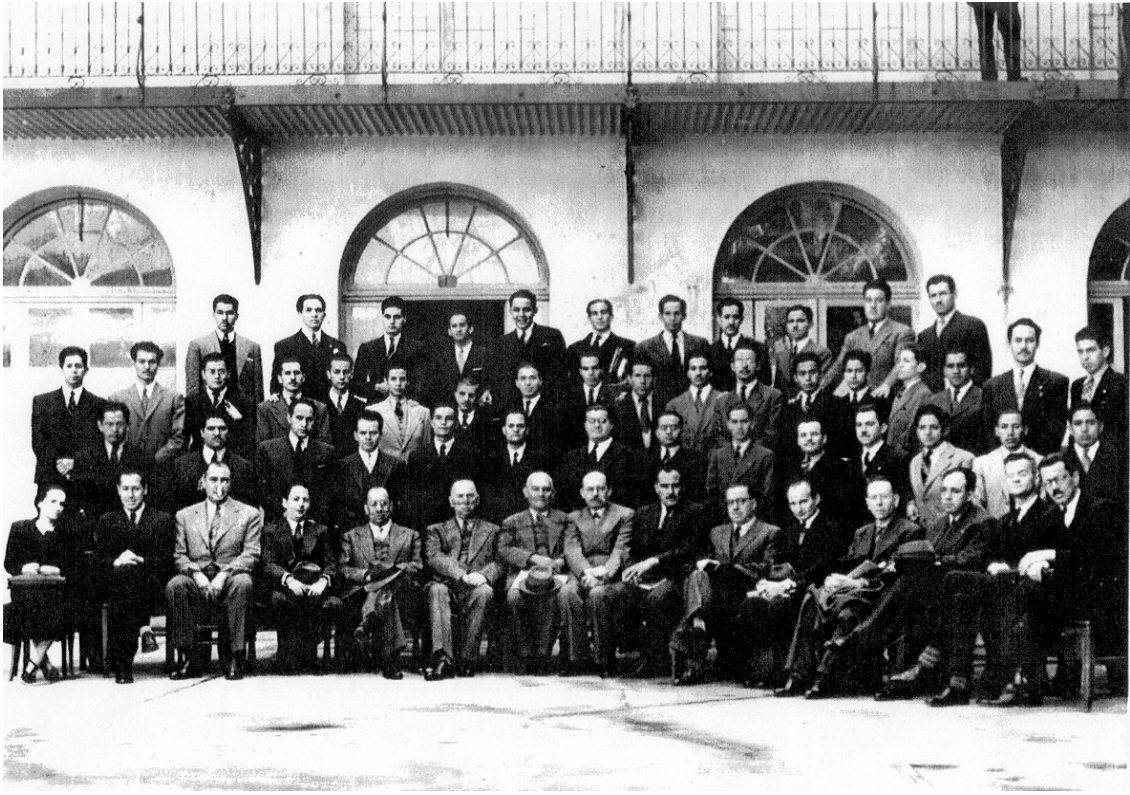

Fig. 8: Teaching staff at the *Escuela Superior de Ingeniería Mecánica y Eléctrica* of the Polytechnical Institute in Mexico City around 1940; Blau in first row on far left. (Archive Keller family)

Like all immigrants from Europe, Blau was not only impressed by the rich Mexican culture but also the view of the beautiful snow-covered volcanoes Popocatépetl and Iztaccíhuatl and enjoyed the pleasant climate. Denied the chance to conduct scientific research, she turned to applied studies, like the effect of solar radiation on the health of the Mexican people,[B44] or the seismic activity in Mexico.[B48]

Austrian intellectuals in Mexico developed a diverse cultural life, organized in the Heinrich Heine Club, but also political activities, centered in the *Acción Republicana Austriaca de México* (ARAM), in which all the political groups cooperated. In 1942, it held a meeting called *Austria Will Rise Again*; moreover, the organization edited a monthly journal *Austria libre*, arranged a weekly broadcast *Voice of Austria*, and regular cultural events. Some of ARAM's leading members were communists, a fact they did not publicize in Mexico. Blau became a member of ARAM.

Also in 1942, Mexico declared war on Germany and Italy.

7.3. U.S.A.

In 1944, about a year after her mother had died, Marietta Blau obtained permission to immigrate into the United States.[13] She moved to New York City, where her younger brother

Ludwig was living, and took jobs in industry, namely, with the research department of the International Rare Metals Refinery Inc. and later with the Canadian Radium and Uranium Company, concentrating on metrology[B53,B55] and industrial applications of radioactivity.[B54] For instance, she published on the use of the photomultiplier tube in conjunction with a scintillating screen to detect radioactive emissions, the first rudimentary scintillation counter,[B49] a great advance over manual counting of light flashes by human observers.[14] Further examples are a radioactive light source,[B50] the measurement of surface areas using a two dimensional α-source,[B52] a radio therapeutic method to reduce the growth and occurrence of tumors.[B51] When in summer of 1947, Blau was – to her dismay – transferred to a small town in Wisconsin, she started writing to colleagues, universities, and firms in the hope of returning to scientific research.

She certainly followed the important discoveries achieved with the photographic method that year, above all the detection of the π meson (pion), with highest interest. At the same time, the plans of the Atomic Energy Commission to construct and apply accelerating machines for nuclear-research work required efficient working groups for the detection of high-energy particles and their reactions: the setup of cloud chambers, bubble chambers, and arrangements of scintillation counters, but most of all the photographic method whose importance had become evident, and for which there were no experts at American universities and the research laboratories created in connection with the Manhattan project.

Employing her in a government program was difficult because she was not yet a U.S. citizen. Finally, she was employed at Columbia University in New York as a scientific staff member as of 1948. Here, Blau was again the only female scientist. Her task was to adopt the photographic emulsions for the study of high-energy particles and to investigate their track properties. For the former, she suggested a two-bath method for the development of thick emulsions.[B56] The latter led to the theoretical derivation of a relationship between grain density in tracks (i. e., number of developed AgBr grains per unit length) and particle energy which is extremely important as it allows to conclude to particle energy from the observed track even if this is not contained in the emulsion as a whole.[B57,B58] Blau's derivation was based on a comparison with ion interactions in strong electrolytes as treated by the theory of Debye and Hückel.

It is quite striking that in these studies Blau methodically reeducated herself on the techniques of experimental emulsion work and acquainted herself with the principal technical achievements since 1938, making significant contributions to the field at the same time.

Although Blau's work at Columbia aimed at the use of emulsions in accelerator experiments (to be performed at Brookhaven National Laboratory), she also exposed plates in balloon flights at high altitudes to gather new experience in evaluating emulsions. She published on a strange event, caused by high-energy cosmic rays.[B59] The star was interpreted as capture of a τ meson (today called kaon) by a bromine or silver nucleus in the emulsion, and raised big attention in these years when the "particle zoo" was established. Another balloon exposure allowed Blau and her co-workers to compare the high-altitude star and meson production and their dependence on material and thickness of absorbers.[B63]

Blau and her colleagues also investigated the intensity measurement of slow neutrons with the photographic method,[B60] comparing the slow-neutron sensitivities of β-sensitive emulsions, x-ray film-indium foil combinations, and boron-loaded plates for epithermal, thermal and cold neutrons. The former two types were found to be equally useful for the detection of epithermal neutrons. $^{10}$B-loaded plates are best for detection of thermal and cold neutrons; very low neutron intensities can be measured by counting of α-tracks.

Blau's construction of a semi-automatic device for analyzing events in nuclear emulsions, together with Sam Lindenbaum and Robert Rudin,[B62,B65] was a landmark work that led not only to future advances in analyzing emulsion tracks, but also portended much later developments in the analysis of bubble chamber, spark chamber, and streamer chamber photographs. Considering the rudimentary level of computers and optical devices in 1950, the capabilities of this device were remarkable. The instrument was built around a microscope with a motor-driven stage which allowed for the photographic plate to be moved in any direction and with any desired speed ($\leq 25$ μm/s). The accuracy of gears, setscrews, etc. was such that dimensions could be measured to within 0.2 μm. A recording chart moved at a speed of 2000 times that of the stage. The image of the plate was observed by the operator through the eyepiece and at the same time was projected on a small slit before a photomultiplier tube. The measurements were done quite rapidly, e. g. the driving time for the grain density record of a 2000 μm track was about ten minutes. The system was adaptable to the measurement of high Z tracks.

In 1950, Marietta Blau was employed at Brookhaven National Laboratory (BNL) which became possible only after she had obtained U.S. citizenship. In the midst of the McCarthy era, individuals suspected of communist collaboration or sympathy were closely scrutinized and investigated. Despite her former membership in the ARAM, in which communist officials had been active, Blau was at last considered politically unobjectionable by the authorities.

At BNL, some of the high-energy machines were already completed. For understanding the interaction of high-energy particles with emulsions as well as for interpreting emulsion experiments, it was required to find out whether the reactions taking place in the emulsion occur on the heavy elements of the silver bromide or the light elements of the gelatin. Blau and collaborators used 300-MeV neutrons[B66] and 50–80-MeV $\pi^+$ mesons[B68] of the Nevis Cyclotron at Columbia University as incident particles and introduced very thin layers of gelatin between photographic emulsion pellicles in order to separate the light and heavy emulsion elements. They determined the percentage of stars from disintegration of light emulsion nuclei, and concluded from mean number and angular distribution of the prongs to reaction mechanisms.

In 1950, when Blau finally had access to the most modern and expensive research techniques, two events that reflected a lack of appreciation for her work before 1938 must have been especially disappointing for her. The Nobel Prize in Physics was awarded to Cecil F. Powell "for his development of the photographic method of studying nuclear processes." Powell had turned to photographic emulsions at Bristol University in 1938. During the war, he established a formidable laboratory and collaboration for the analysis of emulsions and for their improvement by Ilford and Kodak which was achieved by increasing the AgBr content rather than layer thickness. He discovered tracks of π mesons in photographic plates exposed in 1947 at high altitudes in the Bolivian Andes and the Pyrenees, which was partly the reason he was awarded the Nobel Prize. Blau's pioneer work went unrewarded.

In fact, Erwin Schrödinger had proposed Blau and Wambacher for the Nobel Prize in Physics in 1950, pointing out how important their method was in exploring cosmic rays in general and that, in particular they had been the first to interpret the stars correctly as atomic disintegration induced by cosmic rays. Being a Nobel Laureate himself, Schrödinger had counted on the weight of his reputation which did not make up for the lacking support by a lobby, though.

Wambacher had published on *Multiple Disintegration of Atomic Nuclei by Cosmic Rays*[15] already in September 1938, and acquired her *Habilitation* (right of teaching at universities) on the basis of the work *Nuclear Disintegration by Cosmic Rays in Photographic Emulsions*[16] in 1940. After the War, she was removed from the university like all members of the Nazi party, but had been deported to the USSR before. She returned in 1946 and was diagnosed with cancer. When she died in 1950, in several obituaries she was praised for her work as if she had been the preeminent author of the development of the photographic method and the discovery of the disintegration stars. Blau's leading role in this team is insufficiently expressed.

Between 1953 and 1956, Blau and coworkers produced four papers from work at BNL on the interaction of negative pions with emulsion nuclei.[B69–B72] The authors used $\pi^-$ mesons of 500 and 750 MeV from the Brookhaven Cosmotron and identified several events in which two mesons left either an emulsion nucleus or emerged from a hydrogen nucleus, giving unequivocal evidence of meson production. Further events were consistent with additional meson production,[B69] although the tracks were too short for unmistakable identification. Blau and colleagues gave credit to other research groups for finding single events of meson production. Thanks to the higher number of events, it may be stated with considerable justification that Blau's was the first definitive report of additional meson production by high energy mesons, an important, if not unexpected, observation.

Based on the observation of stars with no mesons, stars with one meson and events with two mesons, i. e., additional meson production, Blau and her group performed a careful analysis of meson interactions in nuclear matter.[B71] Interactions with no meson emission (44%) are due to absorption of the incident meson. In events with emission of one meson, the energy and angular distribution suggests a combined process of production, scattering and absorption as possible mechanism of energy loss rather than primary scattering of the incident meson at a nucleon. From the number of stars with two meson emission (meson production), the probability for charged-meson production is derived as 1–3%, and the cross section for the production of charged mesons per nucleon was estimated. Some data on the production of neutral pions were also given. Comparison of the results with those of cosmic-ray mesons, where the mean shower energy is 640 MeV showed a considerable discrepancy, which could not be explained.

In $\pi^-$-experiments with 750 MeV incident energy[B72] also charge exchange played a role, e. g., $\pi^- + p \to \pi^- + \pi^+ + n$ or $\pi^- + p \to \pi^o + \pi^o + n$. Stars without meson emission showed an unexpectedly low number of prongs from which the authors concluded that neutral particles carry away much energy.

Marietta Blau also studied formation and decay of hyperfragments (unstable nuclei with a heavy particle [hyperon[17]] in place of one of the neutrons) and slow K-mesons[B73] detected when a stack of emulsions was exposed to 3-GeV protons (~30,000 protons/cm$^2$), incident parallel to the emulsion layers. Hyperfragments were discovered in 1953; Blau's work was the second systematic investigation of hyperfragments at the Brookhaven Cosmotron, and must be regarded as a pioneering, exploratory effort. A systematic search for hyperfragments in particle beams of well-defined energy gave information on data related to particle physics, production cross section and particle nature of the $\Lambda^o$ hyperon, as well as to physics of the nucleus (formation of the hyperfragment, binding energy, etc.). The individual emulsions of the stack were scanned at low (300×) magnification, since she was searching for fairly prominent stars. Only a few of the events could be analyzed; however, these were compatible with a $\Lambda^o$ hyperon bound to the nucleus. The analysis of the observed stars yielded various nuclear fragments as hyperon carrier, ranging from $^2$H to $^{15}$O. Of the 14,480 stars investigated she found 14 events which were believed to be spontaneous disintegrations of hyperfragments coming to rest in the emulsion. Four stars with emission of $K^+$ mesons were found, with no associated hyperon production.[B73]

Summarizing Blau's work at Brookhaven National Laboratory, it is clear that Marietta Blau stepped authoritatively into the main stream of particle research, in spite of the fact that she was not in command of large research groups. In particular, she quantified the interaction of (then) high energy pion interactions including finding the first examples of additional pion production. She also contributed significantly to the observations of hyperfragments at an early stage. Although the improvement of statistics came several years later with the exploration of hydrogen, deuterium and helium bubble chambers, the path of further research was made clear by the emulsion results.

In 1955, Blau (Fig. 9) became increasingly dissatisfied with her employment at Brookhaven. Not all the projects that were difficult were of interest to her. She felt tired and the atmosphere in the laboratory was not optimal, in fact, quite competitive. The laboratory is situated on the Eastern part of Long Island, New York, at some distance from the Long Island Railroad. Thus Marietta found herself once more in a rural area, isolated from large cities where she could have attended cultural events. She found a place to stay in Patchogue, a village about six miles southeast of the lab. At long last, she received her driver's license and bought a car.

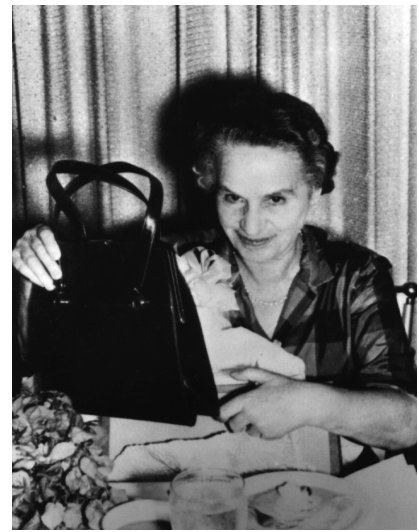

She did not consider retiring from her work in science: The results of new research still fascinated her, and retirement would have left her with too little money as only the years she had worked in the U.S.A. entitled her to a pension. She took a leave of absence from the lab and accepted a position as an associate professor at the University of Miami, Florida, initially for just one semester.

The University of Miami, a private university located in Coral Gables, was chartered in 1925 by a group of citizens when the community grew rapidly during the South Florida land boom and an institution of higher learning was deemed necessary. In the 1940s, a graduate school and the schools of marine science and engineering were added. In the mid-1950s, total enrollment was about eleven thousand. New facilities and resources were aimed at increasing the research productivity of the institution, and doctoral programs were added in many fields.

Fig. 9: Marietta Blau in the 1950s (Archive R. & L. Sexl)

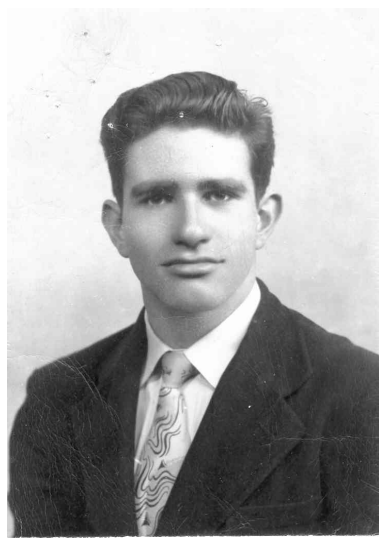

Fig. 10: Arnold Perlmutter around 1960 (Archive Arnold Perlmutter)

In February 1956, Blau drove by herself from New York to Miami. She initially taught various topics related to physics. When she decided to stay beyond one semester, she bought devices and laboratory equipment using funds from U.S. Air Force grants. Because she was so well versed and confident in her abilities in science, she was soon collaborating productively with several colleagues and directed numerous student research projects. She motivated young Arnold Perlmutter (Fig. 10), who had come to Miami shortly before her as a university assistant and who was working in solid-state physics, to join her research program in particle physics. The improved experimental equipment aided the determination of the ionization parameter, the characteristic quantity for particle identification. The particles now being investigated were antiprotons and negative π and K mesons. Collaboration with Perlmutter yielded a series of joint publications. Moreover, Blau became a close friend of Perlmutter and his family; it was as if she were one of them.

Blau built up a laboratory with generous funding from the Air Force Office of Scientific Research (AFOSR), purchasing six precision Leitz binocular microscopes with magnifications up to 2000×, with enlarged movable stages, built by local instrument makers, and a multiple scattering microscope, whose large foundation and massive stage were designed by a master instrument maker near Berkeley, California. In 1958, Blau and her team set up a semi-automatic apparatus, which was inspired by the earlier device and designed here (by Sylvan C. Bloch) with a phototransistor.[18]

Marietta Blau was a most effective teacher, giving courses in electromagnetism and nuclear physics, among others, to advanced undergraduates and graduate students. But she also schooled her collaborators in the laboratory in the theory of ionization measurements, multiple scattering, and range-energy relations. They recruited housewives and students to be scanners of the emulsion pellicles, and in the cases of the more gifted assistants to allow them to make precision measurements.

Blau initiated the evaluation of an emulsion stack (30 layers of 100 μm each) exposed to the $\pi^-$-beam (1.3 GeV/c) of the Brookhaven Cosmotron.[B75] The total length of analyzed tracks was 381 meters, 811 pion interactions were found. A survey of observed interaction processes showed the main reactions to be: $\pi^- + p \to \pi^- + p$ (elastic scattering), $\pi^- + p \to \pi^- + p + \pi^o$ (inelastic scattering), $\pi^- + p \to \pi^- + \pi^+ + n$ (inelastic scattering). These processes occurred on free as well as peripheral protons. Interactions with peripheral neutrons were $\pi^- + n \to \pi^- + n$ (quasielastic scattering) and further events that may have been $\pi^- + n \to \pi^- + n + \pi^o$ or $\pi^- + n \to p + \pi^- + \pi^-$. Considering other investigators' data from cloud and bubble chamber experiments, too, Blau and coauthors calculated values for the pion mean free path and for the cross section of pion-proton collisions and pion interactions with charge transfer. Both the distribution of events of inelastic pion-proton scattering vs. nucleon momentum and angular distribution of emitted nucleons agreed with the isobar model of pion production.[B75]

Blau and her team also evaluated emulsions that had been exposed at the Berkeley Bevatron. Those exposed to $K^+$ mesons of momentum of 620 MeV/c could not be used due to fogging. Further experiments were done with antiprotons of 670 MeV/c[B77] and $K^-$ mesons of 450 MeV/c.[B78,B80] Publication of the latter results concentrated on some extraordinary hyperon interactions, i. e., reactions of hyperons or hyperfragments formed when slowed down $K^-$ mesons are absorbed by nuclei in the emulsion.

Blau vigorously attacked the problems of ionization in nuclear emulsions,[B76] which she had addressed earlier in articles at Columbia University. From data obtained by scanning emulsion tracks of protons, pions and antiprotons with the semi-automatic device it was found that a newly introduced parameter, namely blob density, could be related to the probability of ionization over the entire energy siege. The blob density is the number of developed AgBr single grains or clusters per unit length of track, and replaces the grain density, in this way avoiding uncertainties in counting closely neighboring grains. The quantity is derived from a theoretical expression relating the average probability of development taken over all crystals in the path under consideration, to the value of the ionization, determined by the number of ionization acts caused by the particle on its path.[B76]

At Erwin Schrödinger's request, Blau was awarded the Leibniz Medal of the German Academy of Sciences in Berlin in 1957 for having first observed disintegration stars and for developing the method of detection. As a U.S. citizen, she was not allowed to accept this honor because it originated from the Academy in East Germany, with which the United States did not have diplomatic relations. Even though none of these circumstances changed, she was granted the medal two years later, again at Schrödinger's request, but had to reject it this time as well by order of the State Department. Also in the late 1950s, Schrödinger nominated Blau for the Nobel Prize in Physics again, as had been done equally unsuccessfully in 1955 by the Viennese physicist Hans Thirring

At that time, health problems began to plague Blau, primarily a heart condition and cataracts. Her poor eyesight prevented her from driving. In 1959, she fell and broke her left arm, which required surgery and an extended leave of absence. At that time, she had already decided to return to Vienna. She needed an operation on her eyes, which she was advised not to have done in Florida so shortly after the surgery on her arm and which she could more easily afford in Europe than in the United States. On the other hand, she desired to finish her scientific work in Miami and an article on nuclear emulsions she had agreed to contribute for the series *Methods of Experimental Physics*, edited by her friend, the physicist Chien-Shiung Wu, and her husband, Luke C. L. Yuan.

## 8. Marietta Blau's remaining years in Vienna

Blau finally returned to Vienna in the spring of 1960. After an absence of twenty-two years, one of the first things she experienced in her native country was the fiftieth anniversary of the Radium Institute. She once again became an unpaid worker at the Radium Institute and had to support herself on her pension from the United States, including her medical care, because she did not have health insurance.

Those like Marietta Blau who had been expelled by the Nazi regime were not particularly well received upon their return to Austria. In the 1950s and 1960s, Austria had little interest in the return of emigrants, and generally no serious efforts were made to reintegrate successful researchers into scientific life. Blau never attempted to re-establish her Austrian citizenship which would have required undergoing the same procedure as foreign-born residents.

Meanwhile, the professors that had belonged to the Nazi party had received positions as full professors at the university again. When they were fired at the end of World War II, they had protested against their dismissal, denied their NSDAP membership and wrote extensive justifications. In this way, they got their names removed from the Nazi party registration list and their credentials to teach reinstated.

On the other hand, those who had been victims of the Nazi regime themselves supported Blau: Karl Przibram, together with six other members of the Austrian Academy of Sciences – mainly professors of physics and chemistry at university institutes – proposed her as a corresponding member of the Academy of Sciences, with which they did not succeed, and also nominated her for the Schrödinger Prize which she did receive in 1962 for "the development of the basic photographic method, for the investigation of elementary particles, and in particular for the discovery of disintegration stars together with Dr. Wambacher."

Berta Karlik, now head of the VRI, offered Blau to stay in her small apartment close to the Radium Institute, and put a small room at the institute at her disposal.

Walter Thirring, professor for theoretical physics at the Vienna University, had arranged for Blau to be advisor to a high-energy group consisting of four doctoral students and four additional women who analyzed photographic plates as well as bubble-chamber photographs taken at the European Organization for Nuclear Research, CERN, near Geneva. The plates were scanned by the women, which meant they were searched under microscopes for stars, i.e., for divergent tracks of reaction products of proton-proton interactions. Blau accepted another female doctoral student whose work she supervised from 1960 to 1964. Thirring offered Blau a paid position as leader of this group, but she declined the offer because she feared the obligations might be too demanding for her.

The photographic plates were sent from CERN to the VRI as well as to a group in Bern. Each plate was triple scanned, by three different people, to minimize errors as much as possible. Blau instructed the scanners to send for her immediately when they found certain stars which might indicate hitherto unknown elementary particles. She would then measure these stars herself, despite her poor eyesight. In this laboratory Blau sometimes gave lectures on high-energy experiments for her collaborators, particularly with respect to the discovery of new elementary particles. In summer 1961, Blau (Fig. 11) gave talks at the University of Bern as well as at CERN, which gained considerable attention. She traveled to Switzerland together with two of the young women students of her group.

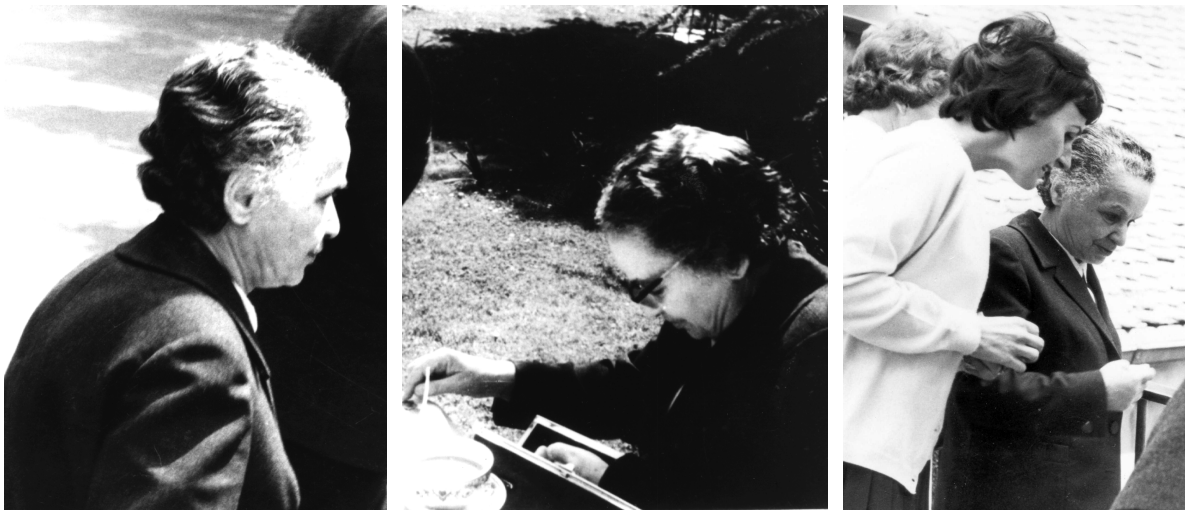

Fig. 11: Marietta Blau during her stay in Geneva, May 1961 (with L. Eggstain; Archive R. & L. Sexl)

Blau completed the chapter on the application of the photographic method in nuclear physics for *Methods of Experimental Physics*,[B79] which was also translated into Russian.[B81]

Blau felt lonesome and disappointed at her lacking reputation, despite her comprehensive scientific achievements. Many of those who knew her considered her withdrawn, and attributed the fact she attracted little attention to her mental disposition of distrust and fear. She even thought of going back to the U.S. in 1962/63. After Blau's last doctoral student had graduated at the end of 1964, Blau quit her activities at the Radium Institute due to her health trouble.

In 1967 Marietta Blau received the Science Prize of the City of Vienna and a plaque from the Paris Radium Institute on the occasion of Marie Curie's hundredth birthday. In 1969, on the occasion of the fiftieth anniversary of Blau's PhD graduation, she could not personally receive the honor of the Golden Doctoral Diploma. Also in 1969, she received letters of recognition from the mayor and the city councilor for culture and public education.

Beginning in 1966, she went to the Lainz hospital for treatments; the diagnoses were sclerosis of the coronary vessels of the heart, pain in her abdomen and problems with her liver and bladder, gall bladder ailment, a tumor in her right lung and suspicion of another in her pelvis. When the lung cancer metastasized, her status became critical and she had to be hospitalized for treatment at the end of September 1969. Her brother Otto came to Vienna to accompany her. Blau died on January 27, 1970. In the obituary notice, Otto characterized her: "Her life was dedicated to science and filled with kindness and charity."

In an announcement of Blau's death in the almanac of the Austrian Academy of Sciences, Karlik wrote: "The Institute regrets the death of an excellent member of long standing, Prof. Dr. Marietta Blau, who passed away in a hospital in Vienna after a long illness on January 27, 1970. Marietta Blau pioneered the method of nuclear emulsions and applied this method with great success, in particular in high-energy physics…"

No scientific journal ever published an obituary for Marietta Blau.

Thirty years after her death, another triumph of the photographic technique demonstrated the importance of her work: The discovery of the τ-neutrino, the last lepton required as proof of the standard model of particle theory, was achieved by use of nuclear emulsion plates in 2000.

In 2004, a memorial tablet honoring Marietta Blau and her scientific work was installed on the Rahlgasse school where she had passed her *Matura* in 1914. Also in 2004, a lecture hall of the university was named after Marietta Blau, and so was a street in Vienna's 22$^{nd}$ district.

## Marietta Blau's scientific publications in chronological order

B1  M. Blau, Über die Absorption divergenter γ-Strahlung, Sitzungsber. Akad. Wiss. Wien, Math. Naturwiss. Kl. IIa 127 (1918) 1253–1279; Mitt. Inst. Radiumf. 110 (1918).

B2  M. Blau, K. Altenburger, Über einige Wirkungen von Strahlen II, Z. Phys. 12 (1922) 315–329.

B3  M. Blau, K. Altenburger, Über eine Methode zur Bestimmung des Streukoeffizienten und des reinen Absorptionskoeffizienten von Röntgenstrahlen, Z. Phys. 25 (1924) 200–214.

B4  M. Blau, Über die Zerfallskonstante von RaA, Sitzungsber. Akad. Wiss. Wien, Math. Naturwiss. Kl. IIa 133 (1924) 17–22; Mitt. Inst. Radiumf. 161 (1924).

B5  M. Blau, Über die photographische Wirkung natürlicher H-Strahlen, Sitzungsber. Akad. Wiss. Wien, Math. Naturwiss. Kl. IIa 134 (1925) 427–436; Mitt. Inst. Radiumf. 179 (1925).

B6  M. Blau, Die photographische Wirkung von H-Strahlen aus Paraffin und Aluminium, Z. Phys. 34 (1925) 285–295.

B7  M. Blau, E. Rona, Ionisation durch H-Strahlen, Sitzungsber. Akad. Wiss. Wien, Math. Naturwiss. Kl. IIa 135 (1926) 573–585; Mitt. Inst. Radiumf. 190 (1926).

B8  M. Blau, Über die photographische Wirkung von H-Strahlen II, Sitzungsber. Akad. Wiss. Wien, Math. Naturwiss. Kl. IIa 136 (1927) 469–480; Mitt. Inst. Radiumf. 208 (1927).

B9  M. Blau, Über die photographische Wirkung von H-Strahlen aus Paraffin und Atomfragmenten, Z. Phys. 48 (1928) 751–764.

B10 M. Blau, Über photographische Intensitätsmessungen von Poloniumpräparaten, Sitzungsber. Akad. Wiss. Wien, Math. Naturwiss. Kl. IIa 137 (1928) 259–268; Mitt. Inst. Radiumf. 220 (1928).

B11 M. Blau, E. Rona, Weitere Beiträge zur Ionisation durch H-Partikeln, Sitzungsber. Akad. Wiss. Wien, Math. Naturwiss. Kl. IIa 138 (1929) 717–731; Mitt. Inst. Radiumf. 241 (1929).

B12 M. Blau, E. Rona, Anwendung der Chamié'schen photographischen Methode zur Prüfung des chemischen Verhaltens von Polonium, Sitzungsber. Akad. Wiss. Wien, Math. Naturwiss. Kl. IIa 139 (1930) 275–279; Mitt. Inst. Radiumf. 257 (1930).

B13  M. Blau, Quantitative Untersuchung der photographischen Wirkung von α- und H-Partikeln, Sitzungsber. Akad. Wiss. Wien, Math. Naturwiss. Kl. IIa 139 (1930) 327–347; Mitt. Inst. Radiumf. 259 (1930).

B14  M. Blau, E. Kara-Michailova, Über die durchdringende Strahlung des Poloniums, Sitzungsber. Akad. Wiss. Wien, Math. Naturwiss. Kl. IIa 140 (1931) 615–622; Mitt. Inst. Radiumf. 283 (1931).

B15  M. Blau, Über photographische Untersuchungen mit radioaktiven Strahlungen, in: F. Dessauer (Ed.), *Zehn Jahre Forschung auf dem physikalisch-medizinischen Grenzgebiet*, Georg Thieme Verlag, Leipzig (1931) 390–398.

B16  M. Blau, Über das Abklingen des latenten Bildes bei Exposition mit α-Partikeln, Sitzungsber. Akad. Wiss. Wien, Math. Naturwiss. Kl. IIa 140 (1931) 623–628; Mitt. Inst. Radiumf. 284 (1931).

B17  M. Blau, H. Wambacher, Über das Verhalten einer kornlosen Emulsion gegenüber α-Partikeln, Sitzungsber. Akad. Wiss. Wien, Math. Naturwiss. Kl. IIb 141 (1932) 467–474; Mitt. Inst. Radiumf. 291 b (1932); also Monatsh. Chem. 61 (1932) 99–106.

B18  M. Blau, H. Wambacher, Über Versuche, durch Neutronen ausgelöste Protonen photographisch nachzuweisen, Anz. Akad. Wiss. Wien 69 (1932) 180–181; Mitt. Inst. Radiumf. 296 a (1932).

B19  M. Blau, H. Wambacher, Über Versuche, durch Neutronen ausgelöste Protonen photographisch nachzuweisen II, Sitzungsber. Akad. Wiss. Wien, Math. Naturwiss. Kl. IIa 141 (1932) 617–620; Mitt. Inst. Radiumf. 299 (1932).

B20  M. Blau, Eine neue Fremdabsorption in Alkalihalogenidkristallen, Nachr. Ges. Wiss. Göttingen II 51 (1933) 401–405.

B21  M. Blau, H. Wambacher, Über den Einfluss des Kornzustands auf die Schwärzungsempfindlichkeit bei Exposition mit α-Partikeln, Z. Wiss. Photogr. Photophys. Photochem. 31 (1933) 243–250.

B22  M. Blau, La méthode photographique et les problèmes de désintégration artificielle des atomes, J. Phys. Radium (Serie 7) 5 (1934) 61–66.

B23  M. Blau, H. Wambacher, Physikalische und chemische Untersuchungen zur Methode des photographischen Nachweises von H-Strahlen, Sitzungsber. Akad. Wiss. Wien, Math. Naturwiss. Kl. IIa 143 (1934) 285–301; Mitt. Inst. Radiumf. 339 (1934).

B24  M. Blau, H. Wambacher, Versuche nach der photographischen Methode über die Zertrümmerung des Aluminiumkernes, Sitzungsber. Akad. Wiss. Wien, Math. Naturwiss. Kl. IIa 143 (1934) 401–410; Mitt. Inst. Radiumf. 344 (1934).

B25  M. Blau, H. Wambacher, Zum Mechanismus der Desensibilisierung photographischer Platten, Z. Wiss. Photogr. Photophys. Photochem. 33 (1934) 191–197.

B26  M. Blau, H. Wambacher, Die photographische Methode in der Atomforschung, Photogr. Korresp. 70, Suppl. 5 (1934) 31–40.

B27  M. Blau, H. Wambacher, Photographic desensitisers and oxygen, Nature (London) 134 (1934) 538.

B28  M. Blau, H. Wambacher, Über die Empfindlichkeit desensibilisierter photographischer Schichten in Abhängigkeit vom Luftsauerstoff und von der Konzentration der Desensibilisatoren, Sitzungsber. Akad. Wiss. Wien, Math. Naturwiss. Kl. IIa 144 (1935) 403–408; Mitt. Inst. Radiumf. 367 (1935).

B29  M. Blau, H. Wambacher, Zum Mechanismus der Desensibilisierung photographischer Platten II, Z. Wiss. Photogr. Photophys. Photochem. 34 (1935) 253–266.

B30  M. Blau, Über den Einfluss des Luftsauerstoffes auf den photographischen Prozess der Ausbleichung, Photogr. Korresp. 71, Suppl. 3 (1935) 21–28.

B31  M. Blau, H. Wambacher, Zur Frage der Verteilung der α-Bahnen der Radiumzerfallsreihe, Sitzungsber. Akad. Wiss. Wien, Math. Naturwiss. Kl. IIa 145 (1936) 605–609; Mitt. Inst. Radiumf. 387 (1936).

B32  M. Blau, H. Wambacher, Über den desensibilisierenden Einfluss von Chlor- und Bromsalzlösungen auf mit Farbstoffen imprägnierte photographische Schichten, Photogr. Korresp. 72 (1936) 108–109.

B33  M. Blau, H. Wambacher, Bemerkungen zur Desensibilisierungstheorie von K. Weber, Z. Wiss. Photogr. Photophys. Photochem. 35 (1936) 211–215.

B34  M. Blau, H. Wambacher, Längenmessung von H-Strahlbahnen mit der photographischen Methode, Sitzungsber. Akad. Wiss. Wien, Math. Naturwiss. Kl. IIa 146 (1937) 259–272; Mitt. Inst. Radiumf. 397 (1937).

B35  M. Blau, H. Wambacher, Vorläufiger Bericht über photographische Ultrastrahlenuntersuchungen nebst einigen Versuchen über die 'spontane Neutronenemission'. Auftreten von H-Strahlen ähnlichen Bahnen entsprechend mehreren Metern Reichweite in Luft, Sitzungsber. Akad. Wiss. Wien, Math. Naturwiss. Kl. IIa 146 (1937) 469–477; Mitt. Inst. Radiumf. 404 (1937).

B36  M. Blau, H. Wambacher, Disintegration processes by cosmic rays with the simultaneous emission of several heavy particles, Nature (London) 140 (1937) 585.

B37  M. Blau (after joint experiments with H. Wambacher), Über die schweren Teilchen in der Ultrastrahlung (abstract of lecture), Verh. der Deutschen Phys. Ges. 18 (1937) 123.

B38  M. Blau, H. Wambacher, II. Mitteilung über photographische Untersuchungen der schweren Teilchen in der kosmischen Strahlung. Einzelbahnen und Zertrümmerungssterne, Sitzungsber. Akad. Wiss. Wien, Math. Naturwiss. Kl. IIa 146 (1937) 623–641; Mitt. Inst. Radiumf. 409 (1937).

B39  M. Blau, Photographic tracks from cosmic rays, Nature (London) 142 (1938) 613.

B40  M. Blau, H. Wambacher, Die photographische Methode in der Atomforschung. II. Bericht, Photogr. Korresp. 74 (1938) 2–6 and 23–29.

B41  M. Blau, Über das Vorkommen von Alpha-Teilchen mit Reichweiten zwischen 1.2 und 2 cm in einer Samariumlösung, Arch. Math. Naturvidensk. B 42 (4) (1939) 1–10.

B42  M. Blau, Sobre la existencia de una radiación α cuyo origen hasta ahora se desconoce, Ingeniería (Mexico) 14 (1940) 50–56.

B43  M. Blau, El helio. Su origen y su localización, Ciencia (Mexico) 1 (1940) 265–270.

B44  M. Blau, La radiación solar en las condiciones de México, Ciencia (Mexico) 3 (1942) 149–157.

B45  M. Blau, Investigation of the radioactivity of rocks and thermal springs in Mexico, Yearbook Am. Phil. Soc. 1943, 134–135.

B46  M. Blau, Algunas investigaciones sobre radiactividad llevadas a cabo en México, Ciencia (Mexico) 5 (1944) 12–17.

B47  M. Blau, Notas para la medición de pequeñas corrientes de ionización, Rev. Mex. Electr., April 1944, 5–7.

B48  M. Blau, La radiactividad y el estado térmico de la tierra, Ciencia (Mexico) 5 (1944) 97–103.

B49  M. Blau, B. Dreyfus, The multiplier photo-tube in radioactive measurements, Rev. Sci. Instrum. 16 (1945) 245–248.

B50  M. Blau, I. Feuer, Radioactive light sources, J. Opt. Soc. Am. 36 (1946) 576–580.

B51  M. Blau, H. Sinason, O. Baudisch, Radioactivation of colloidal gamma ferric oxide, Science 103 (1946) 744–748.

B52  M. Blau, J. Carlin, Ionization currents from extended alpha-sources, Rev. Sci. Instrum. 18 (1947) 715–721.

B53  M. Blau, H. Sinason, Routine analysis of the alpha activity of protactinium samples, Science 106 (1947) 400–401.

B54  M. Blau, J.R. Carlin, Industrial applications of radioactivity, Electronics 21 (1948) 78–82.

B55  M. Blau, J.E. Smith, Beta-ray measurements and units, Nucleonics 2 (6) (1948) 67–74.

B56  M. Blau, J.A. De Felice, Development of thick emulsions by a two-bath method, Phys. Rev. 74 (1948) 1198.

B57  M. Blau, Grain density in photographic tracks of heavy particles, Phys. Rev. 75 (1949) 279–282.

B58  M. Blau, Grain density in nuclear tracks (abstract of lecture), Phys. Rev. 75 (1949) 1327.

B59  M. Blau, M.M. Block, J.E. Nafe, Heavy particles in cosmic-ray stars, Phys. Rev. 76 (1949) 860–861.

B60  M. Blau, I.W. Ruderman, J. Czechowski, Photographic methods of measuring slow neutron intensities, Rev. Sci. Instrum. 21 (1950) 232–236.

B61 M. Blau, Bericht über die Entdeckung der durch kosmische Strahlung erzeugten 'Sterne' in photographischen Emulsionen, Sitzungsber. Österr. Akad. Wiss., Math. Naturwiss. Kl. IIa 159 (1950) 53–57.

B62 R. Rudin, M. Blau, S. Lindenbaum, A semi-automatic device for analyzing events in nuclear emulsions (abstract of lecture), Phys. Rev. 78 (1950) 319–320.

B63 M. Blau, J. Nafe, H. Bramson, The dependance of high altitude star and meson production rates on absorbers (abstract of lecture), Phys. Rev. 78 (1950) 320.

B64 M. Blau, Möglichkeiten und Grenzen der photographischen Methode in Kernphysik und kosmischer Strahlung, Acta Phys. Austriaca 3 (1950) 384–395.

B65 M. Blau, R. Rudin, S. Lindenbaum, Semi-automatic device for analyzing events in nuclear emulsions, Rev. Sci. Instrum. 21 (1950) 978–985.

B66 M. Blau, A.R. Oliver, Stars induced by high energy neutrons in the light elements of the photographic emulsion (abstract of lecture), Phys. Rev. 87 (1952) 182.

B67 M. Blau, E.O. Salant, T-Tracks in nuclear emulsions, Phys. Rev. 88 (1952) 954–955.

B68 M. Blau, A.R. Oliver, J.E. Smith, Neutron and meson stars induced in the light elements of the emulsion, Phys. Rev. 91 (1953) 949–957.

B69 M. Blau, M. Caulton, J.E. Smith, Meson production by 500-MeV negative pions, Phys. Rev. 92 (1953) 516–517.

B70 M. Caulton, M. Blau, J.E. Smith, Interactions of 500-MeV negative pions with emulsion nuclei (abstract of lecture), Phys. Rev. 93 (1954) 919.

B71 M. Blau, M. Caulton, Inelastic scattering of 500-MeV negative pions in emulsion nuclei, Phys. Rev. 96 (1954) 150–160.

B72 M. Blau, A.R. Oliver, Interaction of 750-MeV $\pi^-$ mesons with emulsion nuclei, Phys. Rev. 102 (1956) 489–494.

B73 M. Blau, Hyperfragments and slow K-mesons in stars produced by 3-BeV protons, Phys. Rev. 102 (1956) 495–501.

B74 M. Blau, Ionisationsmessungen in photographischen Emulsionen, Acta Phys. Austriaca 12 (4) (1959) 336–355.

B75 M. Blau, C.F. Carter, A. Perlmutter, Negative pion interactions at 1.3 GeV/c, Nuovo Cimento 14 (1959) 704–721; also PB Report 145040, USDCTS (1959); also Department of Defense Report AFOSR-TN-59-788 (1959).

B76 M. Blau, S.C. Bloch, C.F. Carter, A. Perlmutter, Studies of ionization parameters in nuclear emulsions, Rev. Sci. Instrum. 31 (1960) 289–297; also Department of Defense Report AFOSR-TN-59-229 (1959).

B77 M. Blau, C.F. Carter, A. Perlmutter, Study of antiproton interactions, Department of Defense Report AFOSR-TN-60-461 (1960).

B78 M. Blau, C.F. Carter, A. Perlmutter, Interaction and decays of hyperons produced in K-capture stars at rest, PB Report 149322, USDCTS (1960); also Department of Defense Report AFOSR-TN-60-745 (1960).

B79 M. Blau, in: L.C.L. Yuan and C.-S. Wu (Eds.), *Methods of Experimental Physics – Vol 5: Nuclear Physics*, Academic Press, New York and London (1961, 1963).

  (1) Section 1.7. Photographic emulsions, Vol. 5A, 208–264.
  (2) Section 2.1.1.3. Charge determination of particles in photographic emulsions, Vol. 5A, 298–307.
  (3) Section 2.2.1.1.5. Momentum measurement in nuclear emulsions, Vol. 5A, 388–408.
  (4) Section 2.2.3.8. Detection and measurement of gamma-rays in photographic emulsions, Vol. 5A, 676–682.
  (5) Section 2.3.5. Determination of mass of nucleons in emulsions, Vol. 5B, 37–44.

B80 M. Blau, C.F. Carter, A. Perlmutter, An example of hyperfragment decay in the $\pi^+$ mode and other interactions of $K^-$ mesons and hyperons in emulsion, Nuovo Cimento 27 (1963) 774–785.

B81 M. Blau, Fotografitseckie emulsii, in: *Principy i metody registracii elementarnich tschastits*, Moscow (1963) (partial translation of contributions in *Methods of Experimental Physics – Vol. 5: Nuclear Physics*).

---

[1] Viktor F. Hess, Über Beobachtungen der durchdringenden Strahlung bei sieben Freiballonfahrten, Phys. Z. 13 (1912) 1084–1091.

[2] Fritz Paneth, Georg v. Hevesy, Über Radioelemente als Indikatoren in der analytischen Chemie, Sitzungsber. math.-nat. Kl. IIa 122 (1913) 1001; Mitt. Inst. Radiumf. 43 (1913).

[3] E. Rutherford, Collision of α particles with light atoms. IV. An anomalous effect in nitrogen, Phil. Mag. Ser. 6 37 (1919) 581–587.

[4] Wilhelm Michl, Zur photographischen Wirkung der α-Teilchen, Sitzungsber. Akad. Wiss. Wien, Math. Naturwiss. Kl. IIa 123 (1914) 1955; Mitt. Inst. Radiumf. 68 (1914).

[5] Hertha Wambacher, Untersuchung der photographischen Wirkung radioaktiver Strahlungen auf mit Chromsäure und Pinakryptolgelb vorbehandelte Filme und Platten, Sitzungsber. Akad. Wiss. Wien, Math. Naturwiss. Kl. IIa 140 (1931) 271; Mitt. Inst. Radiumf. 274 (1931).

[6] James Chadwick, On the possible existence of a neutron, Nature 129 (1932) 312.

[7] With α-particles from polonium incident on beryllium, this is called Po-Be neutron source.

[8] W. Bothe, W. Kolhörster, Das Wesen der Höhenstrahlung, Z. Phys. 56 (1929) 751–777.

[9] Arthur H. Compton, Nature of Cosmic Rays, Nature 131 (1933) 713–715.

[10] K. von Meyenn (Hg.), Wolfgang Pauli – Wissenschaftlicher Briefwechsel mit Bohr, Einstein, Heisenberg u. a., Springer-Verlag, Berlin (1985), vol. II, p. 495, vol. III, p. 7, 9, 29, 31.

[11] W. Heisenberg, Der Durchgang sehr energiereicher Korpuskeln durch den Atomkern, Naturwissenschaften 25 (1937) 749–750; also Ber. Sächs. Akad. Wiss., math.-phys. Kl. 89 (1937) 369–384.

[12] H. Euler, W. Heisenberg, Theoretische Gesichtspunkte zur Deutung der kosmischen Strahlung, Ergebn. exakt. Naturwiss. 17 (1938) 1.

[13] The report on Blau's research in the U.S. is based on Arnold Perlmutter's article "Marietta Blau's Work After World War II" (in English) contributed to the original (German) book: Robert Rosner, Brigitte Strohmaier (Hg.), Marietta Blau – Sterne der Zertrümmerung, Biographie einer Wegbereiterin der modernen Teilchenphysik, Böhlau, Wien (2003).

[14] Further developments, using liquid, plastic and crystal scintillators, soon made the scintillation counter a pre-eminent detector in nuclear and particle physics.

[15] Hertha Wambacher, Mehrfachzertrümmerung durch kosmische Strahlung; Ergebnisse aus 154 Zertrümmerungssternen in photographischen Platten, Phys. Z. 39 (1938) 883.

[16] Hertha Wambacher, Kernzertrümmerung durch Höhenstrahlung in der photographischen Emulsion, Sitzungsber. Akad. Wiss. Wien, Math. Naturwiss. Kl. IIa 149 (1940) 157; Mitt. Inst. Radiumf. 435 (1940).

[17] Hyperons are elementary particles belonging to the class of baryons, which means they consist of three quarks and possess half integer spin and strangeness. They decay via weak interaction. Baryon types well known in the 1950s were proton and neutron.

[18] S.C. Bloch, Gap length analyzer for nuclear emulsion tracks, Rev. Sci. Instr. 29 (1958) 789–790.